\documentclass[twocolumn,showpacs,preprintnumbers,amsmath,amssymb]{revtex4}
\usepackage{graphicx}
\begin{document}

\title{Study of radiation background at various high altitude locations in preparation for rare event search in cosmic rays}

\author{R. Bhattacharyya$^a$, S. Dey$^a$, Sanjay K. Ghosh $^{a,b}$, A. Maulik$^a$\footnote{Corresponding author. Fax: +91 33 25693125.\\E-mail address: atanu.maulik@jcbose.ac.in (Atanu Maulik)}, Sibaji Raha$^{a,b}$, D. Syam$^a$}

\affiliation{$^a$Centre for Astroparticle Physics and Space Science,~Bose Institute, Kolkata 700 091, India}
\affiliation{$^b$Department of Physics,~Bose Institute, Kolkata 700 009, India}

\date{\today}

\begin{abstract}
Various phenomenological models presented over the years have hinted at the possible presence of strangelets, which are nuggets of Strange Quark Matter (SQM), in cosmic rays. One way to search for such rare events is through the deployment of large area Nuclear Track Detector (NTD) arrays at high mountain altitudes. Before the deployment of any such array can begin, a detailed study of the radiation background is essential. Also, a proper understanding of the response of detectors exposed to extreme weather conditions is necessary. With that aim, pilot studies were carried out at various high altitude locations in India such as Darjeeling (2200 m a.m.s.l), Ooty (2200 m a.m.s.l) and Hanle (4500 m a.m.s.l). Small arrays of CR-39 as well as high threshold Polyethylene Terephthalate (PET) detectors were given open air exposures for periods ranging from three months to two years. The findings of such studies are reported in this paper.  
\end{abstract}

\maketitle

\section{Introduction}

Various phenomenological models developed over the years have raised the possibility that nuggets of Strange Quark Matter (SQM), called strangelets, are present in cosmic rays ~\cite{madsen,sayan,sayan2,sayan3}. Such strangelets will have nearly equal numbers of up, down and strange quarks and so will have an anomalous $Z/A$ ratio ($Z/A \ll 1/2$), where $Z$ is the charge and $A$ is the baryon number, compared to ordinary nuclei. Some models~\cite{banerjee,wu} of strangelet propagation through the earth's atmosphere have strongly hinted at the possible presence of low energy ($\sim$~few MeV/n) strangelets at high mountain altitudes, although with a very low flux. Also, over the years, various experimental groups have reported the presence of particles with anomalous $Z/A$ ratios in cosmic rays~\cite{saito,ichimura}. But none of those groups have been able to make any definitive claim because of lack of statistics. So the search for strangelets in cosmic rays remains an active area of research~\cite{prl}.

One of the best ways to look for low energy strangelets with very low fluxes at high mountain altitudes is through the deployment of very large area arrays of Nuclear Track Detectors (NTDs). Such passive detector arrays offer several advantages over many other detector types. They are relatively inexpensive to deploy, easy to maintain and do not require any power for their operation, a fact which offers particular advantages when it comes to the deployment of large arrays at very remote high altitude locations. Also because of their high intrinsic thresholds of registration, some NTDs offer a natural way to suppress the huge low-$Z$ background (Neutron recoil tracks, atmospheric radon alpha tracks etc.) expected in any such experiment. 

Nuclear Track Detectors (NTDs) like CR-39, Makrofol etc. have been used for charged particle detection for many years ~\cite{FL75,DU87}. We plan to use a particular brand of commercially available polymer, identified as Polyethylene Terephthalate (PET)~\cite{pet}, as NTD, in the search for exotic particles in cosmic rays, through the deployment of large area arrays of PET films at high mountain altitudes. PET was found~\cite{basu,nimb16} to have a much higher detection threshold ($Z/\beta\sim 140$, where $Z$ is the charge and $\beta=v/c$ the measure of the velocity of the impinging particle) compared to other commercially available NTDs like CR-39, Makrofol etc. ($Z/\beta$ threshold lying in the range 6-60). This makes PET particularly suitable for low energy rare event search in cosmic rays.  

Before any new material can be employed as a detector, it needs to be properly characterized and calibrated. With that aim, systematic studies were carried out on PET to determine its ideal etching condition and also to ascertain its charge response to various ions using accelerators as well as natural radioactive sources. A calibration curve for PET ($dE/dx$ vs. $V_T/V_B$, where $dE/dx$ is the specific energy loss, $V_T$ and $V_B$ are the track and the bulk etch rates respectively while their ratio $V_T/V_B$ is the reduced etch rate or charge response) utilizing $^{16}$O, $^{32}$S, $^{56}$Fe, $^{238}$U ions was obtained~\cite{nimb}. It was then updated~\cite{astro} with additional data points corresponding to $^{129}$Xe, $^{78}$Kr and $^{49}$Ti ions obtained from the REX-ISOLDE facility at CERN. Also, studies were carried out to determine the charge and energy resolution that could be achieved with PET~\cite{nimb14}. All these studies have firmly established PET as a very efficient detector of heavily ionizing particles with a detection threshold much higher than the other commercially available detector material CR-39 which is in widespread use today. 

In addition to such calibration experiments, pilot studies were carried out at different high altitude locations where PET, as well as standard CR-39 detectors, were given open air exposures for durations ranging from a few months to two years. The goal was to study how the detector behaviour changes with exposure to harsh environmental conditions and also to survey the local radiation background. The results of such studies are presented in this paper. 
  
\section{Exposures at high mountain altitudes}

The sites chosen for these studies are Darjeeling in Eastern Himalayas, Hanle in Northern Himalayas and Ooty in the Nilgiri Hills. Table~\ref{table:parameters} lists the altitudes (above mean sea level) and some other parameters associated with those sites. One reason for the choice of these sites is the existence of scientific research facilities there, which is going to make the eventual deployment and maintenance of any such large area array easier. Also these facilities have records of climatic conditions at those sites going back many years and they continue to collect and maintain such records.

\begin{table*}[ht]
{{\caption{\label{table:parameters}Some parameters of the sites where NTDs were given open air exposures}}
\begin{tabular}{|c|c|c|c|c|c|c|} \hline
Place       & Altitude   & Mean         & Atmospheric   & Geographic                            & Geomagnetic    &Geomagnetic \\
            & (a.m.s.l)  & Atmospheric  & Depth         & latitudes and                         & latitudes and  &cutoff       \\
            &            & Pressure     &               & longitudes                            & longitudes     &rigidity    \\ \hline 
            & (m)        & (hPa)        & $(gcm^{-2})$  &                                       &                &(GV)           \\ \hline
Darjeeling  & $2200$ & $780$ & $795$   & $27.0$$^{\circ}$N, $88.3$$^{\circ}$E  & $17.6$$^{\circ}$N, $162.2$$^{\circ}$E & $14.7$ \\ \hline
Ooty        & $2200$ & $785$ & $800$   & $11.4$$^{\circ}$N, $76.7$$^{\circ}$E  & $ 2.9$$^{\circ}$N, $149.9$$^{\circ}$E & $15.9$  \\ \hline
Hanle       & $4500$ & $580$ & $591$   & $32.8$$^{\circ}$N, $78.9$$^{\circ}$E  & $24.1$$^{\circ}$N, $154.1$$^{\circ}$E & $14.2$   \\ \hline
\end{tabular}}
\end{table*} 

\begin{figure}[ht]
\centering
\includegraphics[width=0.8\hsize,clip]{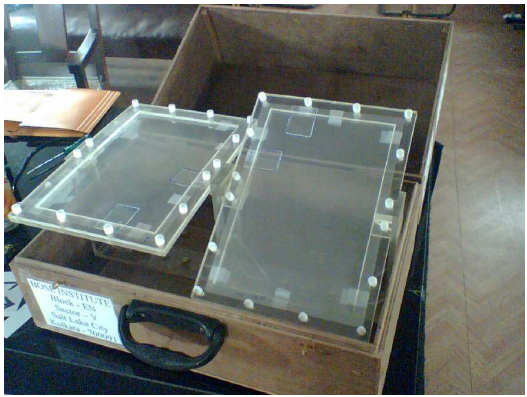}
\caption{\label{detectors} A4 size PET films as well as smaller pieces of CR-39 mounted on perspex stands.} 
\end{figure}

Stacks containing three PET films of A4 size ($297~mm \times 210~mm$) and thickness $90~\mu m$ as well as CR-39 films of size $5~cm \times 5~cm$ and thickness $700~\mu m$ were mounted on perspex stands and given open air exposures for durations ranging from three months to two years at those sites. Fig~\ref{detectors} shows PET films as well as smaller pieces of CR-39 mounted on perspex stands. The stands in turn are fitted inside a box for convenience of transport.   

\section{Results and discussions}

After the exposures, the plastic films were brought back to the laboratory for analysis. The PET and CR-39 films were etched for durations ranging from 1--4 h in 6.25 N NaOH solution at 55.0$\pm$0.5$^{\circ}$C and 70.0$\pm$0.5$^{\circ}$C respectively, which were found \cite{nimb} to be the ideal etching condition for these plastics. 

The PET films were found to have retained their flexibility. Even those films exposed at Hanle, where the temperature drops down to -30$^{\circ}$C during winter, did not show any embrittlement. But the bulk etch rates for the detector films were found to change after prolonged exposures. This change is especially significant for CR-39, whereas PET is found to be relatively more robust. Table~\ref{table:vb} lists the bulk etch rates ($V_B$) of some PET and CR-39 detector films after open air exposures for various durations, as well as that of unexposed films for comparison. 

\begin{figure}[ht]
\centering
\includegraphics[width=0.8\hsize,clip]{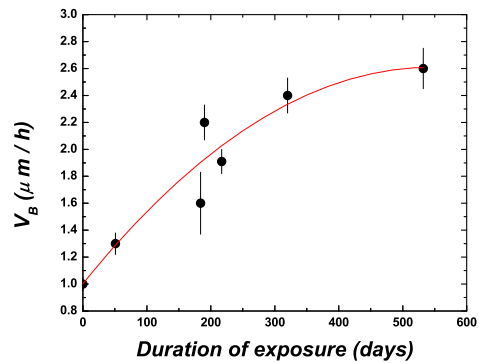}
\caption{\label{vbvsd} Plot showing the variation in the bulk etch rates with the duration of exposure of PET detector films} 
\end{figure}

Fig~\ref{vbvsd} shows how the bulk etch rates ($V_B$) of PET films increase with the increase in the duration of exposure. This increase in the bulk etch rates with exposure is expected, as the chemical bonds near the top surface of the uppermost detector film in a stack can break due to exposure to UV rays and other weather elements, thereby enhancing the rate of etching for those portions. But the saturation of the rate of etching also indicates that the damage is mostly restricted to the top few microns of the film, which are etched out in the first hours of etching. So we do not expect this change to affect our overall analysis. Nevertheless, we are considering limiting the exposure duration for one set of detectors to a maximum of few months and placing the detectors in protective packaging in order to limit the changes in bulk etch rates with exposure.
 
\begin{table}[ht]
{{\caption{\label{table:vb}The bulk etch rates of CR-39 and PET}}
\begin{tabular}{|c|c|c|c|} \hline
Place       & Exposure & $V_B$ of                    & $V_B$ of               \\
            & duration & CR-39                       & PET                     \\
            &          &                             &                          \\ \hline 
            & (Days)   & ($\mu$m/h)                  &($\mu$m/h)                 \\ \hline
Darjeeling  & 532      & 11.7$\pm$0.7                & 2.6$\pm$0.15               \\ \hline
Ooty        & 190      & 10.0$\pm$0.6                & 2.2$\pm$0.13                \\ \hline
Hanle       & 320      & 11.5$\pm$0.7                & 2.4$\pm$0.13                 \\ \hline
Unexposed   & --       & 1.4$\pm$0.07                & 1.0$\pm$0.05                  \\ \hline
\end{tabular}}
\end{table}

\begin{figure}[ht]
\centering
\includegraphics[width=0.8\hsize,clip]{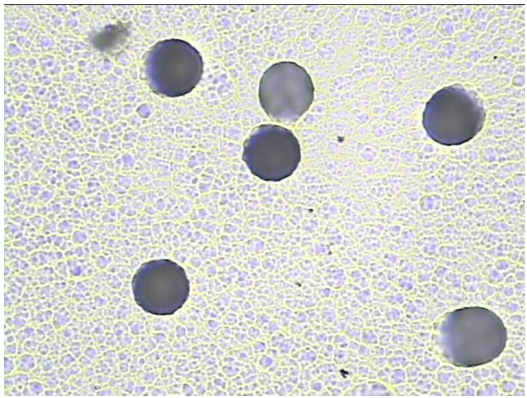}
\caption{\label{darjeeling} Charged particle tracks seen on CR-39 exposed at Darjeeling after 4 h etching. The size of the image frame is 117 $\mu$m $\times$ 87 $\mu$m.} 
\end{figure}

The etched samples were analyzed with Leica optical microscopes interfaced with computers for image analysis. Fig~\ref{darjeeling} shows some tracks observed on CR-39 film exposed at Darjeeling, after the sample has been etched for 4 h. 

The particle flux measured on CR-39 and PET at the three sites are given in Table~\ref{table:flux}. An analysis of PET films exposed at Ooty and Hanle, as well as the latest batch of PET films exposed at Darjeeling did not reveal any charged particle tracks allowing us to set limits. This is in line with expectations, as PET detectors, with higher detection threshold compared to CR-39, does not record proton or alpha particle tracks and heavier ions are not expected to be present in any appreciable quantity in cosmic rays at these depths in the atmosphere. 

\begin{table}[ht]
{{\caption{\label{table:flux}Flux recorded on CR-39 and PET at the three sites}}
\begin{tabular}{|c|c|c|} \hline
Place       & Flux on                    & Flux on    \\
            & CR-39                      & PET         \\
            &                            &              \\ \hline 
            & $(cm^{-2}s^{-1}sr^{-1})$ & $(cm^{-2}s^{-1}sr^{-1})$  \\ \hline
Darjeeling  & $6.0 \times 10^{-4}$       & $< 1.0 \times 10^{-11}$      \\ \hline
Ooty        & $1.3 \times 10^{-4}$       & $< 4.1 \times 10^{-11}$       \\ \hline
Hanle       & $4.6 \times 10^{-4}$       & $< 2.4 \times 10^{-11}$        \\ \hline
\end{tabular}}
\end{table} 
             
\begin{figure}[ht]
\centering
\includegraphics[width=0.8\hsize,clip]{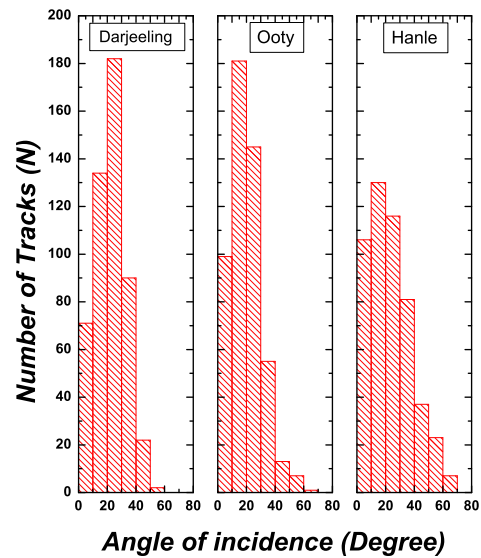}
\caption{\label{angle}Distribution of the angle of incidence of  tracks recorded on CR-39 at Darjeeling, Ooty and Hanle.} 
\end{figure}

\begin{figure}[ht]
\centering
\includegraphics[width=0.8\hsize,clip]{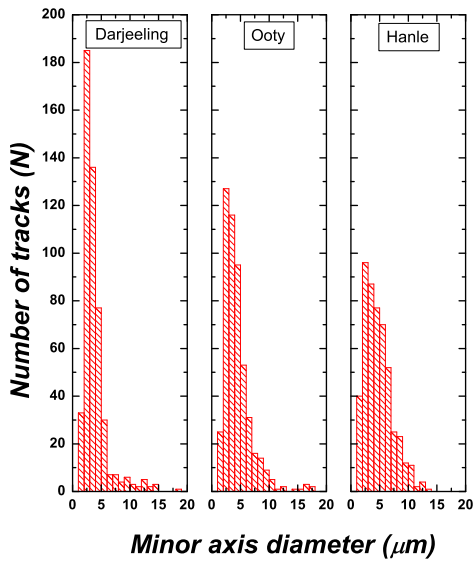}
\caption{\label{diameter}Distribution of the minor axis diameter of tracks recorded on CR-39 at Darjeeling, Ooty and Hanle (Samples etched for 4 h).} 
\end{figure}

\begin{figure}[ht]
\centering
\includegraphics[width=0.8\hsize,clip]{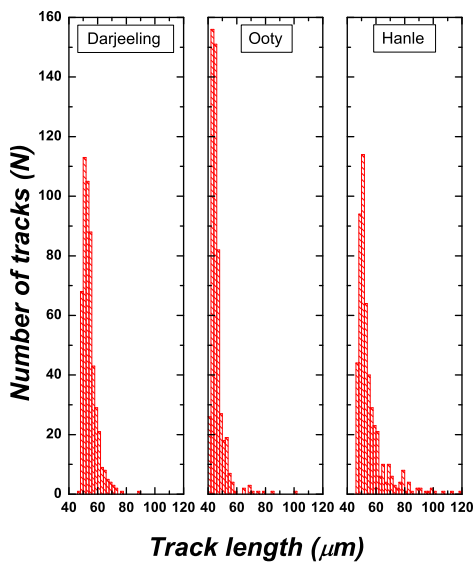}
\caption{\label{tracklength}Distribution of the length of tracks recorded on CR-39 at Darjeeling, Ooty and Hanle (Samples etched for 4 h).} 
\end{figure}

\begin{figure}[ht]
\centering
\includegraphics[width=0.8\hsize,clip]{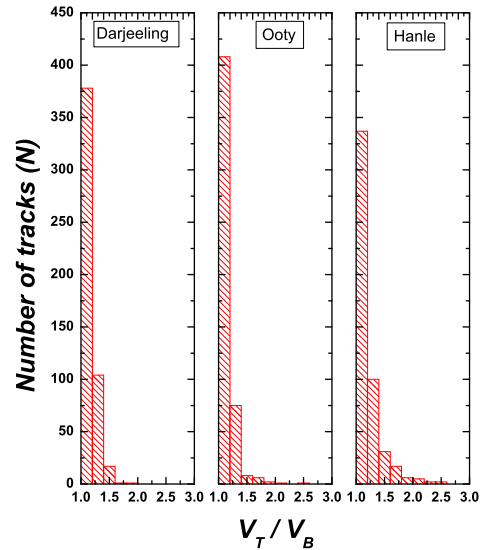}
\caption{\label{vtvb}Distribution of the $V_T/V_B$ values of tracks recorded on CR-39 at Darjeeling, Ooty and Hanle.} 
\end{figure}

Fig.~\ref{angle}, Fig.~\ref{diameter}, Fig.~\ref{tracklength}, Fig.~\ref{vtvb} give distribution of the angle of incidence, minor axis diameter, track length and $V_T/V_B$ values for the tracks recorded on CR-39 at Darjeeling, Ooty and Hanle. 

We plan to set up an array of PET detectors covering an area of 100 sq.m. for a total exposure duration of 1 yr (if necessary through multiple deployments of shorter durations). This will allow us to reach a sensitivity of $\sim 10^{-13}cm^{-2}s^{-1}sr^{-1}$ and put to the test competing models of strangelet propagation, as even with the most conservative assumptions, the model due to Banerjee et.al.~\cite{banastro}, which posits that the mass of a strangelet will grow during its passage through the atmosphere, predicts a strangelet flux in excess of $\sim 10^{-12}cm^{-2}s^{-1}sr^{-1}$ at high mountain altitudes($\sim$~3000~m). In sharp contrast, the propagation model of Wilk et. al.~\cite{wilk,wilk2} conjectures that once inside the atmosphere, the mass of even an initially large strangelet will decrease rapidly through collisions with air molecules and eventually the strangelet will evaporate into neutrons. Under this scenario, no significant flux of strangelets is expected at mountain altitudes. So a non-observation of any track over a 100 sq.m. array will strongly dis-favor the Banerjee model. 

Since heavy ion fluxes from primary cosmic rays at mountain altitudes ($\sim$~3000~m) are estimated to be $\sim 10^{-14}cm^{-2}s^{-1}sr^{-1}$~\cite{grieder}, we do not expect primary cosmic ray heavy ions to constitute any significant background in strangelet search with PET detectors. We would argue that the use of PET films as NTDs, by virtue of their high detection threshold and hence their ability to suppress the entirety of low-Z background, will allow us to look for even very low energy strangelets ($\sim$~few MeV/n) which may stop in the first layer of a NTD stack. This enables us to probe a unique region of parameter space which is inaccessible to other conventional NTDs like CR-39 or to any active detector.    

\section{Conclusion}
From the studies conducted so far, high threshold PET seems to be a very good choice as a NTD for the planned rare event search.  We will continue to conduct further studies along these lines in order to select a particular site as well as begin the process of large scale search for strangelets using large area NTD arrays.

\begin{acknowledgments}
The authors sincerely thank staff members of IIA at Hanle and of TIFR at Ooty for their help in setting up the detectors at those places. The authors also thank Mr. Sujit K. Basu and Mr. Deokumar Rai of Bose Institute for technical assistance. The work is funded by IRHPA (Intensification of Research in High Priority Areas) Project (IR/S2/PF-01/2011 dated 26.06.2012) of the Science and Engineering Research Council (SERC), DST, Government of India, New Delhi.
\end{acknowledgments}

\end{document}